\documentclass[aps,nofootinbib,showpacs]{revtex4}

\usepackage{graphicx}
\usepackage{colordvi}

\newcommand{\lsim}[1]{
\setlength{\unitlength}{12pt}
\begin{picture}(1.4,1.)
\put(.7,-0.3){\makebox(0.0,1.)[t]{$<$}}
\put(.7,-0.3){\makebox(0.0,1.)[b]{$\sim$}}
\end{picture}#1}
\newcommand{\gsim}[2]{
\setlength{\unitlength}{12pt}
\begin{picture}(1.4,1.)
\put(.7,-0.3){\makebox(0.0,1.)[t]{$>$}}
\put(.7,-0.3){\makebox(0.0,1.)[b]{$\sim$}}
\end{picture}#2}
\begin{document}

\title{Constraining theories of low-scale quantum gravity by non-observation
of the bulk vector boson signal from the Sun}
\author{R. Horvat, D. Kekez\thanks{%
Corresponding author. \newline
\textit{E-mail address}: kekez@irb.hr (D. Kekez).}, Z. Kre\v{c}ak and A.
Ljubi\v{c}i\'{c}}
\affiliation{Rudjer Bo\v{s}kovi\'{c} Institute, P.O.B. 180, 10002 Zagreb, Croatia}

\begin{abstract}
In this experiment we aim to detect Kaluza-Klein (KK)
excitations of the bulk gauge field,
emitted in a bremsstrahlung process on solar plasma constituents, by looking
at a process analogous to the photoelectric effect inside the HPGe detector.
Using a generic feature of the underlying effective theory that the unknown
4-dimensional gauge coupling is independent of the number of extra \textit{%
large} dimensions $\delta $, we show that the expected number of events in
the detector is insensitive to the true scale of quantum gravity for $\delta
= 2$. With the entire data collection time of 202 days in the energy
interval 1.7 - 3.8 keV, the number of events detected was as low as $%
1.1\cdot 10^{6}$, compared to $2.7\cdot 10^{6}$ from the expected high
multiplicity of the solar KK excitations for $\delta =2$. Hence, our bound
from the presumed existence of new forces associated with additional gauge
bosons actually conforms with very stringent bounds set from various
astrophysical considerations.
Baring any modifications of the infrared part of the KK spectrum,
this bound would therefore rule out the possibility of
observing any signal at the LHC for $\delta = 2$. Although a dependence on
the fundamental scale referring to $(4 + \delta )$-dimensional gravity turns
on again for $\delta = 3$, the experimental sensitivity of the present setup
proves insufficient to draw any constraint for $\delta > 2$.
\end{abstract}

\maketitle




In a remarkable proposal of Arkani-Hamed, Dimopoulos and Dvali (ADD) \cite{1}%
, the scenario of large extra dimensions has been advocated as an
alternative viewpoint capable to shed some light on the gauge hierarchy
problem . The radical revision from the standard description of physics
beyond SM lies in the fact that low-scale (right above the weak scale)
Planck mass $M_*$ and Newton's constant $M_{Pl}^{-2}$ do peacefully coexist,
if gravity is allowed to propagate in $\delta$ extra \textit{large } compact
dimensions. By applying Gauss' law, the simple case of the original
Kaluza-Klein (KK) theory \cite{2} can be generalized to $\delta $ extra
dimensions as 
\begin{equation}
M_{Pl}^{2} = M_{*}^{2 +\delta } V_{\delta } \;,
\end{equation}
where, considering a $\delta $-dimensional torus of radius $R$, $V_{\delta }
= (2 \pi R)^{\delta }$.

Although the ADD proposal is considered only as a reformulation of the
hierarchy problem and not a full solution \footnote{%
Instead of explaining why the Higgs VEV is small, one now needs to explain
why $V_{\delta }$ is huge compared with $M_{*}^{-\delta }$. Moreover, in
this scenario the hierarchy problem becomes closely related to the
cosmological constant problem \cite{3}.}, it is still of non-ceasing
interest nowadays due to its remarkable phenomenological implications \cite%
{4}. Besides bulk gravitons, the same scenario, in which heavy mass scales
in four dimensions are replaced by lighter mass scales in higher dimensions,
can be applied to other non-SM fields as well. Examples cover successful
application to neutrino \cite{5} as well as axion phenomenology \cite{6}.
Since the SM has been tested down to distances $10^{-16}$ cm, much shorter
than the compactification radius R, such class of higher-dimensional
theories is nowadays conventionally considered in the context of the brane
paradigm, where all the SM degrees of freedom are localized on a 3-brane.
The realization of the ADD proposal is possible in type I string theory \cite%
{7}, for example, with the SM degrees of freedom localized on a D-brane.

On the observational side, there are two main classes of laboratory tests
for the ADD proposal. The first searches for possible deviations from
Newton's inverse-squared law in table top experiments \cite{8}. Typically,
we have $M_* > 2$ TeV for $\delta = 2$, although model-dependent effects can
substantially affect these predictions. The second class deals with the
collider bounds, giving $M_* > 1.45 $ (0.65) TeV for $\delta = 2 $ (6) for
the combined LEP-Tevatron data \cite{9}, with an extra possibility to study
the so-called transplanckian regime at the LHC \cite{10}. Much more
stringent bounds on gravity in flat extra dimensions comes from variety of
astrophysical considerations. The most restrictive limits on large extra
dimensions comes from low measured luminosity of some pulsars, giving $M_* >
750 $ (35) TeV for $\delta = 2 $ (3) \cite{11}. However, contrary to the
collider bounds, these bounds are quite sensitive to the infrared part of
the KK graviton spectrum.

Yet another possibility is to have gauge fields in the bulk, gauging $B-L$
(or either $B$ or $L$ separately), for example \cite{4}. The limited impact
of additional gauge bosons at low energy, due to decoupling of
heavy-particle species, may not be the case any more if the new bosons are
allowed to propagate freely in the bulk. The one-loop correction of a bulk
gauge boson to the muon magnetic moment has already been calculated \cite{12}%
. In the present paper, we aim to observe gauge-boson KK excitations coming
from the Sun and emitted in a bremsstrahlung process, $e+X\to e^\prime +X+b$%
, where X is a proton or $\alpha$ particle and ``$b$'' stands for the KK
gauge boson. Our experimental setup has been designed such as to capture KK
modes from the Sun in the HPGe detector, in a process analogous to the
ordinary photoelectric effect, $b+A\to e+A^\prime$, where now $A(A^\prime)$
refers to the germanium atom. Due to the expected high multiplicity of the
solar KK modes, we expect a clear signal above background for $\delta = 2$.
The important point in testing higher-dimensional effective theory lies in
the fact that the relevant combination of the 4-dimensional coupling $g_4 $
and the total number of KK modes in the kinematically allowed interval $%
N_{tot}$, $g_{4}^4 N_{tot}$, turns out to be independent of $M_* $ for $%
\delta = 2$ (see below).
Assuming no modifications were made on the infrared part
of the gauge-boson KK spectrum,
this means that non-observation of the KK signal
would rule out 2 extra dimensions. Unfortunately, our present experimental
setup has no potential to catch a signal above the background for $\delta >
2 $ modes.

Considering U(1) gauge field propagating in the bulk, the effective
4-dimensional coupling, representing a universal coupling of each KK mode to
fermions, can be estimated as \cite{4, 12} as 
\begin{equation}
g_{4}^2 \setlength{\unitlength}{12pt} \begin{picture}(1.4,1.)
\put(.7,-0.3){\makebox(0.0,1.)[t]{$>$}}
\put(.7,-0.3){\makebox(0.0,1.)[b]{$\sim$}} \end{picture}\frac{M_{*}^2 }{%
M_{Pl}^2 } \;,
\end{equation}
being thus independent of $\delta $ for a given $M_{*}$. Taking $M_{*} \sim
1 $ TeV, we get $g_{4} \sim 10^{-16}$, and therefore it was argued in \cite%
{4} that a gauge boson could mediate forces at submillimeter distances $10^6
- 10^8$ times stronger than gravity. On the other hand, the zero mode of a
new gauge boson (coupled to baryons) can not remain massless since masses
larger than $\sim 10^{-4}$ eV are obligatory in order to comply with fifth
force experiments. This means that the gauge symmetry is spontaneously
broken in the bulk \cite{4} or on a distant wall \cite{12} by the VEV $v$ of
some scalar field. Either case gives $m_{A^0 } = g_4 v$. Since the effective
theory is valid up to $M_* $, $g_4 M_* $ represents at the same time the
upper bound to the mass of the zero mode. By plugging (2), we get $m_{A^0 }
\sim R_{\delta = 2}^{-1} \sim 10^{-4}$ eV, and considering the kinematical
cutoff (essentially given by thermal energies of electrons in the Sun of
order $\sim $ keV), we see that only the far-infrared part of the KK
spectrum is affected by $m_{A^0 }$. So we can safely disregard $m_{A^0 }$ in
our considerations.
\footnote{Actually, with the present scenario limits 
on $M_{\star }$ more restrictive than those in \cite{11} can be inferred from
strong astrophysical constraints on the emission of new gauge bosons from
stars \cite{Grifols:1986fc,Grifols:1988fv} (for a review see \cite{Raffelt:2000kp}). One can
easily switch a typical bound on the fine
structure constant of forces coupled to electron number, $\alpha_E \lsim
10^{-28}$ \cite{Grifols:1986fc,Grifols:1988fv}, to a bound on the fundamental scale by
making use of Eq. (2). One obtains, $M_{\star } \gsim \; 10^4$ TeV. In
contrast, our bound, conditioned by both emission and detection
processes, always restricts a quantity $g_{4}^4 N_{tot}$ (differently from
the emission bound which always restrict $g_{4}^2 N_{tot}$), which is
independent of the fundamental scale for $\delta =2$. Hence our experiment 
has a potential to rule out the two extra dimensions altogether (see below).}

With the above preliminaries, we are ready to write down the expected number
of event in the detector, differential with regard to the gauge boson energy 
$E$, as 
\begin{equation}
\frac{dN_b}{dE} = \frac{d \Phi_b}{dE} \sigma_{b + A \to e + A^\prime} N_{%
\mbox{\rm\scriptsize Ge}} t \;,
\end{equation}
where $N_{\mbox{\rm\scriptsize Ge}}$ is the number of germanium atoms in the
detector, and $t$ is the data collection time.

\begin{figure}[tp]
\begin{center}
\includegraphics[width=10cm,angle=0]{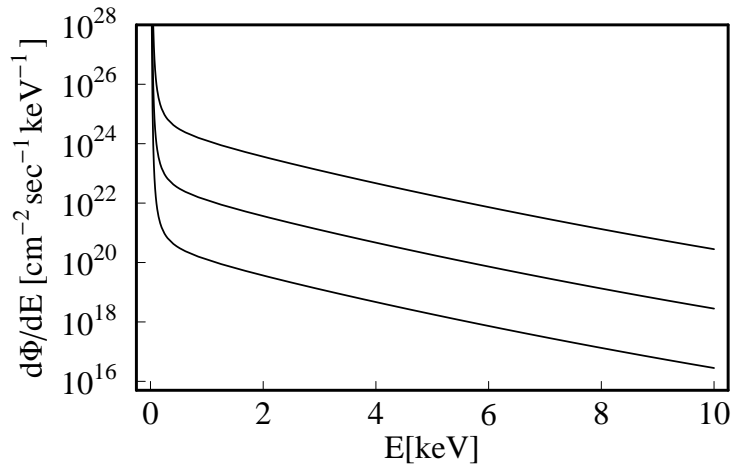}
\end{center}
\caption{Energy dependence of the solar flux of KK gauge bosons at Earth,
for three values of $M_{*}$: 1 TeV, 10 TeV, and 100 TeV (upper,
middle, and lower curve, respectively).
}
\label{fig:SolarFlux}
\end{figure}

The differential flux of gauge bosons at Earth in (3), integrated over a
standard solar model \cite{13}, can be found to be 
\begin{eqnarray}
\frac{d\Phi _{b}}{dE} &=&\frac{1}{d_{\odot }^{2}}\int_{%
\mbox{\rm\scriptsize
Sun}}(n_{H}+4\,n_{\mbox{\rm\scriptsize He}})\,r^{2}\,dr\int_{0}^{+\infty
}dT_{e}\,\frac{d\sigma _{e+p\rightarrow e^{\prime }+p+b}(T_{e})}{dE} 
\nonumber \\
&&\times \sqrt{\frac{2T_{e}}{m_{e}}}\left[ \frac{2\sqrt{T_{e}}n_{e}\beta
^{3/2}\exp (-\beta T_{e})}{\sqrt{\pi }}\right] \;,
\end{eqnarray}%
where $d\sigma _{e+p\rightarrow e^{\prime }+p+b}(T_{e})/dE$ represents a
differential cross section for the bremsstrahlung process in the Sun, and is
given by 
\begin{equation}
\frac{d\sigma _{e+p\rightarrow e^{\prime }+p+b}(T_{e})}{dE}=\frac{\alpha _{4}%
}{\pi }\sigma _{\mbox{\rm\scriptsize Th}}\,\frac{m_{e}}{T_{e}}\,\frac{1}{E}%
\ln \left[ \frac{(\sqrt{T_{e}}+\sqrt{T_{e}-E})^{2}}{E}\right] N_{%
\mbox{\rm\scriptsize tot}}(T_{e}),  \label{BremsstrahlungCrossSection}
\end{equation}%
where $\alpha _{4}=g_{4}^{2}/(4\pi )$. This is just the cross section for $%
e+p\rightarrow e^{\prime }+p+\gamma $ process, known from QED, with the
change $\alpha \rightarrow \alpha _{4}$. As for Eqs.(3-5), a few comments
are in order. Firstly, in order to calculate the production rate of gauge
bosons in a thermal background analytically, we approximate the Fermi-Dirac
electron distribution by a Boltzmann one, i.e., we neglect +1 in front of $%
e^{E_{e}/T}$, what is perfectly justified because the electron mass is much
larger than the solar temperature. Second, $N_{tot}(T_{e})$ used in (\ref%
{BremsstrahlungCrossSection}) captures the effect of massive KK modes, and
represents the number of KK modes whose momentum along the extra dimensions
is less than the electron kinetic energy $T_{e}$ (see e. g. \cite{14}), 
\begin{equation}
N_{tot}(T_{e})=\frac{S_{\delta -1}}{\delta }\frac{M_{Pl}^{2}}{M_{\ast
}^{\delta +2}}T_{e}^{\delta }\;,
\end{equation}%
where $S_{\delta -1}=(2\pi )^{\delta /2}\Gamma (\delta /2)$ is is the
surface of an $\delta $ dimensional sphere with unit radius. We should
stress here that we intentionally make use of this approximation (instead of
calculating cross section for the production of an individual mode with mass 
$m$, and then integrating over $m$'s) in order to show manifestly that by
Eqs.(2) and (6) 
\begin{equation}
g_{4}^{4}N_{tot}(T_{e})\sim T_{e}^{\delta }\frac{M_{\ast }^{2-\delta }}{%
M_{Pl}^{2}}
\end{equation}%
is independent of $M_{\ast }$ for $\delta =2$. Further, $\sigma _{%
\mbox{\rm\scriptsize Th}}=6.65\cdot 10^{-25}\,\mbox{\rm\ cm}^{2}$ denotes
the Thomson scattering cross section, $d_{\odot }$ is the distance to the
Sun while $n_{e}$, $n_{H}$ and $n_{He}$ represent the number densities of
electrons, H and He nuclei in the Sun, respectively.
In Fig.~\ref{fig:SolarFlux}  we
display the flux of KK gauge-bosons at Earth as a function of their energy
using some typical values for $M_{\ast }$.

  In a bremsstrahlung process in the Sun most of the gauge bosons will be
emitted in the low energy region, as is evident from Fig.~\ref{fig:SolarFlux}.
For energies of
a few keV we estimate that the most sensitive process for detection of gauge
bosons will be the boson-electric effect on L, M and N electrons,
respectively, of germanium atoms. Therefore the cross section in (3) is
calculated from photoabsorption cross section $\sigma _{\gamma
+A\rightarrow e+A^{\prime }}$ as
\begin{equation}
\sigma _{b+A\rightarrow e+A^{\prime }}=\frac{\alpha _{4}}{\alpha }\,\sigma
_{\gamma +A\rightarrow e+A^{\prime }~,}
\end{equation}%
\noindent with data for $\sigma _{\gamma +A\rightarrow
e+A^{\prime }}$ taken from Ref.~\cite{NIST}.

\begin{figure}[tp]
\begin{center}
\includegraphics[width=11cm,angle=0]{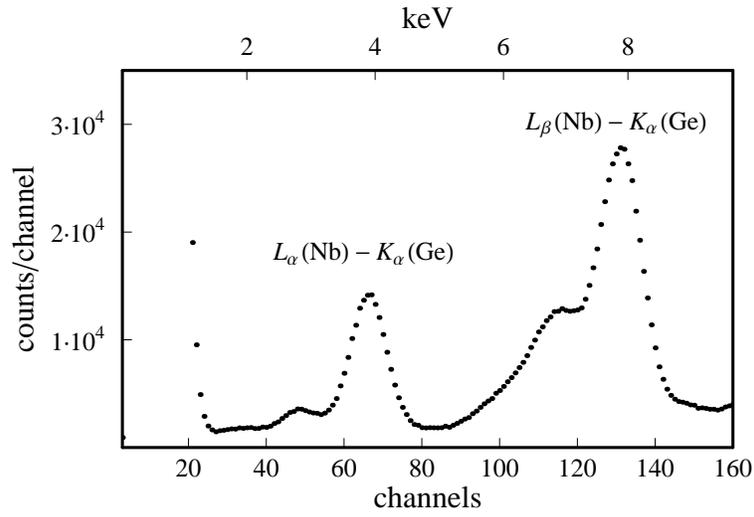}
\end{center}
\caption{Low-energy $^{241}$Am spectra.}
\label{fig:Am241}
\end{figure}

\begin{figure}[bp]
\begin{center}
\includegraphics[width=11cm,angle=0]{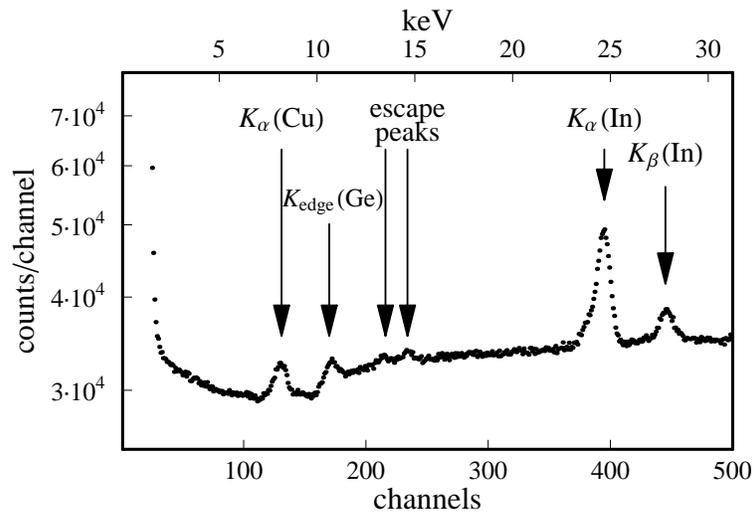}
\end{center}
\caption{Energy spectra accumulated in our HPGe detector for 202 days.}
\label{fig:spectra}
\end{figure}

\begin{figure}[tp]
\begin{center}
\includegraphics[width=11cm,angle=0]{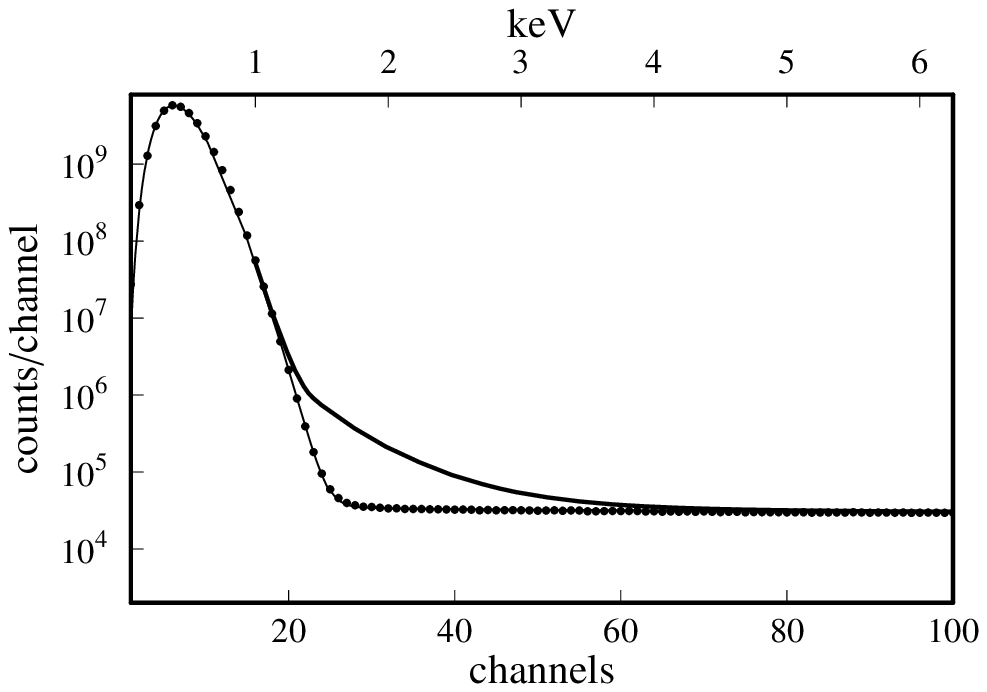}
\end{center}

\caption{Number of events in our detector in different energy channels. The
expected contribution of KK gauge bosons
(thick line), the experimental data (full circles) and fit
through the experimental data points (thin line) are shown.}
\label{fig:spectraLow}
\end{figure}

After being ejected from the germanium atoms, their energies, together with
subsequently emitted X-rays, will be completely absorbed in our large 1.5 kg
HPGe crystal. Detector efficiency
relative to 3"$\times $3" NaI crystal at 30 cm was 65\%, peak-to-Compton
ratio at 1.33 MeV was 60:1, and the FWHM at 5.9 keV was $\sim $800 eV.
Energy calibration was obtained with a set of calibrated radioactive
sources. Possible instabilities were controlled on a daily basis, and were
less than $\pm $1 channel, which is equivalent to $\pm $0.0634 keV. Detector
was placed inside an iron box with internal dimensions 54$\times $33$\times $%
33 cm$^{3}$ and with wall thickness ranging from 16 to 23 cm. The iron was
more than 70 years old and was essentially free of $^{60}$Co impurities. The
box was lined outside with 1 cm thick lead and background events were
reduced by a factor of $\sim $5. Because most of the gauge bosons are
expected to be emitted in the low energy region, special care was taken to
check the HPGe detector behaviour for energies below 10 keV.
The two prominent escape peaks, obtained with radioactive $^{241}$Am source,
can be seen in Fig.~\ref{fig:Am241}. The 3.9 keV peak,
originating from the escape of germanium K$_{\alpha}$
after absorption of niobium L$_{\alpha }$ X-rays,
clearly shows up at 66$^{\mbox{\scriptsize th}}$ channel,
and by fitting procedure we find FWHM$\simeq$663 eV.

  The energy spectra below 30 keV accumulated for 202 days are
shown with full circles in Fig.~\ref{fig:spectra}.
Above a huge electronic noise, which
dominates for energies $<$1.7 keV, there are several visible
peaks. Indium K$_{\alpha }$ and K$_{\beta }$ peaks\ show
presence of indium which serves as (i) vacuum gasket of ORTEC PopTop
detector capsule and (ii) material for making germanium transistors of
preamplifier mounted inside the detector capsule. Two escape peaks are due
to the escape of germanium K$_{\alpha }$ and K$_{\beta }$
X-rays, respectively, after indium K$_{\alpha }$ X-rays are absorbed
in the crystal. The lowest energy peak belongs to copper 8.04 keV
K$_{\alpha }$ X-rays, originating from the copper stick which serves as
holder/cooler of the germanium crystal. The most prominent peak,
indium K$_{\alpha }$, served for monitoring of daily spectra stability. All of
the above features, during whole measurement, show integrity and stability of
total experimental spectrum; there are no systematic errors, i.e. there are
only statistical errors.

 In Fig.~\ref{fig:spectraLow}  the first 100 channels
of Fig.~\ref{fig:spectra}  are shown in more details.
The full thick line represents the expected KK gauge boson events
superimposed on the accumulated energy spectra. The full thin line, which
represents fit through the experimental points, is a sum of electronic
noise and an almost flat background.
Electronic noise is described with a function
$y(k)=1.1\cdot 10^{7}\exp [-0.86(k-0.09)^{2}$k$^{0.92}]k^{6.65}$
and flat part of the spectra with straight line
$y_{b}(k)=(33400-31k)$; $k$ is the channel number. Straight line was
obtained as the best fit through the experimental data in the 60-100
channels interval, where contribution of KK gauge bosons is expected to be
negligible.

 In the energy interval 1.7-3.8 keV the expected number of KK gauge
bosons was $2.7\cdot 10^{6},$ while for the same energy interval
only $1.1\cdot 10^{6}$ events were detected. At $\sim$1.7
keV the ratio of expected KK and detected events was approximately 10:1.
Given the smallness of $g_{4}$, and the highest mass of the KK excitations
in the keV range, in explaining our finding we can safely ignore the
possibility of strong attenuation of the gauge bosons in the Sun as well as
the possibility that some particles do not survive the journey from the
Sun. This means that when this experimental result is interpreted in the
context of higher-dimensional low-scale gravity and the ADD proposal, two
extra dimensions are definitely ruled out, since by Eq. (7) the effect is
expected not to depend on the fundamental scale $M_{\ast }$. Our result
actually supersedes the existing stringent astrophysical bounds for $\delta
=2$, which had already been considered to exclude $\delta =2$, if the ADD
proposal had something to do with the hierarchy problem.

Finally, we explore a potential of our detector to observe $\delta =3$ KK
modes. In this case we see from (7) that a dependence on the fundamental
scale shows up, but the effect is additionally suppressed by a factor of $%
T_e /M_{*}$ compared with the $\delta =2$ case. In the same time/energy
interval as above we have found the expected number of events to be several
orders of magnitude below the background. Since this is far below the
background, we conclude that the present experimental setup has no potential
to explore $\delta >2$ extra dimensions.

In conclusion, we have used a HPGe detector to detect electrons and gammas
produced in a U(1)-electric effect when the presumed massive KK modes of the
bulk vector field hit the detector from the Sun. Our calculation of the
expected number of events assumes that massive KK modes were produced in a
bremsstrahlung process in the Sun. With no signal detected above background,
and a theoretical prediction that for two extra dimensions the expected
number of events in the detector is practically independent of the
fundamental scale of gravity, we can definitely rule out the said number of
extra dimensions. With the existence of new gauge bosons our result
therefore conclusively confirms what a variety of (more model-dependent and
thus more uncertain) astrophysical bounds had indicated earlier. In
searching for the KK signal from more than two dimensions, we conclude that
the limited experimental sensitivity of the present setup has no potential
to uncover this possibility.

\textbf{Acknowledgments. } The authors acknowledge the support of the
Croatian MSES project No.~098-0982887-2872.

\end{document}